\documentstyle[12pt]{article}
\input tcilatex

\QQQ{Language}{
American English
}

\begin{document}

\baselineskip=0.6cm \vspace{1.5cm}

\noindent
{\Large PHYSICS AND CONSCIOUSNESS }

{\normalsize \vspace{1.5cm} }{\large \noindent
Patricio P\'{e}rez }

{\large \vspace{1cm} \noindent
Departamento de F\'{\i}sica, Universidad de Santiago de Chile }

{\large \vspace{0.5cm} \noindent
Casilla 307, Correo 2, Santiago, Chile }

{\large \vspace{1.5cm} ABSTRACT \vspace{1cm} }

{\normalsize Some contributions of physics towards the understanding of
consciousness are described. As recent relevant models, associative memory
neural networks are mentioned. It is shown that consciousness and quantum
physics share some properties. Two existing quantum models are discussed. }

{\normalsize \vspace{1.5cm} \noindent
{\large INTRODUCTION } }

{\normalsize {\large \vspace{1cm} }A possible definition for human
consciousness is that it is the perception of our own mental states. If we
accept that mental states are correlated with physical states in the brain,
then a scientific study of the phenomenon of consciousness should be based
on an understanding of the properties of this complex organ. Especially
during the second half of this century, physics has made important
contributions in order to explain significant aspects of the functioning of
the brain. A sample of this is the modeling of the mechanism for the
propagation of a nerve pulse. We could also mention the application of
statistical mechanics to the simulation of associative memory in the brain.
Furthermore, quantum physics, which assigns a crucial role to the observer,
suggests some possible routes towards the understanding of conscious
phenomena. In this paper we describe in some detail these contributions of
the physical science. \vspace{1.2cm}  }

{\normalsize {\large \noindent
PHYSICS IN THE BRAIN } }

{\normalsize {\large \vspace{1cm} }A basic aspect of the functioning of the
brain is the permanent exchange of nerve pulses between its specialized
cells (neurons). With the help of physics we have now a good understanding
of the mechanisms of generation and propagation of these nerve pulses. Pulse
generation is possible thanks to ion exchange, mainly potassium and sodium
through the neuron's membrane. In the resting state of the neuron there is a
potential difference across the membrane that surrounds not only the body of
the cell but also the branch like structures called {\it axon} and {\it %
dendrites}. The potential is such that it is negative in the interior as
compared with the exterior. Under appropriate conditions a depolarizing
localized stimulus will induce transient ionic currents through the
membrane, which will change the sign of the potential difference. The
creation and relaxation of this perturbation or {\it action potential} is
successfully modeled with a simple electric circuit which has adjustable and
time dependent resistances [1]. The propagation of the action potential
along the axon may be represented by the solution of a wave equation.  }

{\normalsize It is widely acknowledged that the memory capacity of a person
is an essential ingredient in conscious perception. During the last fifteen
years there has been an explosive multidisciplinary interest on models of
neural networks, which among other virtues make possible quantitative
modeling of associative memory. One of the most important associative memory
neural network models is that proposed in 1982 by the physicist John J.
Hopfield [2]. The basic ingredients of this model are the following:  }

{\normalsize The network is defined by a set of simple processing units, all
connected with each other. The state of the network at any instant of time
is determined by the collection of the states of the processing units (or
neurons), which could change from an initial configuration to a final stable
state. The processing units, which mimic the neurons in the brain can be in
any of two states, firing a signal ($+1$) or inactive ($-1$). The state of a
given unit is assigned according to the states of the units connected to it
and to the strength of each of these connections. These connections or
weights are a simple model of the synapses between real neurons. Assuming a
discrete representation of time, the state of neuron $i$ at $t+1$ is
calculated as a function of the state of the other neurons at $t$ as
follows:  }

{\normalsize
\begin{equation}
s_{i}(t+1) = sign \left( \sum_{i} w_{ij} s_{j}(t) \right)
\end{equation}
}

{\normalsize \noindent
where the updating is performed randomly or sequentially, one neuron at a
time. $w_{ij}$ is the weight of the connection between neuron $j$ and neuron
$i$, and $s_{j}(t)$ is the state of neuron $j$ at time $t$. The function $%
sign(x)$ gives a $+1$ whenever $x$ is positive and a $-1$ when $x$ is
negative. Equation (1) may be interpreted as a a dynamical law which will
govern the evolution of the network from any initial state to a final
stationary configuration.  }

{\normalsize Several properties of the Hopfield model may be obtained
exploiting its isomorphism with a spin glass, a system which has been
extensively studied by physicists using the tools of statistical mechanics.
An important parameter, the {\it storage capacity} ($\alpha_{c}$), which
measures the ratio between the maximum amount of stationary configurations
and the total amount of neurons in the network, may be calculated
analytically for different choices of the weights.  }

{\normalsize One possibility for the assignation of the weights is based on
the {\it Hebb rule} [3], which establishes that whenever a a couple of
neurons that are connected are simultaneously active, their synapsis is
strengthened. An implementation of this rule has been studied by Hopfield
and many other scientists. In this case $\alpha_{c}=0.144$.  }

{\normalsize The reasons why a neural network model as that described is
interesting as an associative memory model are:  }

{\normalsize \noindent
-The final stationary configurations of the network may be identified with
concepts memorized by a living being.  }

{\normalsize \noindent
-Synapses are modified through learning, which seems a well established fact
among biologists.  }

{\normalsize \noindent
-Initial states of the network may be interpreted as stimuli presented to
the living being, and the corresponding stationary states reached after
applying the dynamical law may be seen as the concepts associated with the
stimuli.  }

{\normalsize \vspace{1.2cm}  }

{\normalsize {\large \noindent
QUANTUM PHYSICS AND CONSCIOUSNESS } }

{\normalsize {\large \vspace{1cm} }Quantum physics assigns an essential role
to the observer of an event or experiment. Classical physics instead rests
on the assumption that there exists an objective reality, which is
independent of wether somebody is scrutinizing it or not. The relation
quantum event - observer (assuming that quantum effects are important for
our understanding of the properties of the brain) may lead us to think that
quantum physics will explain consciousness. Let us consider for example the
following words said by the philosopher J. R. Searle in his recent book
''The rediscovery of the mind'' [4]:  }

\begin{quotation}
{\normalsize ''consciousness is not reducible in the way other phenomena are
reducible, not because the pattern of facts in the real world involves
anything special, but because the reduction of other phenomena depend in
part on distinguishing between 'objective physical reality', on the one
hand, and mere 'subjective appearance, on the other; and eliminating the
appearance from the phenomena that have been reduced. But in the case of
consciousness, its reality is the appearance; hence, the point of the
reduction would be lost if we tried to carve off the appearance and simply
defined consciousness in terms of the underlying physical reality''.  }
\end{quotation}

{\normalsize Referring to the wave function that describes the state of a
quantum system, the physicist W. Heisenberg says [5]:  }

\begin{quotation}
{\normalsize ''The observation itself changes the probability function
discontinuously; it selects of all possible events the actual one that has
taken place... the transition from the 'possible' to the 'actual' takes
place during the act of observation. If we want to describe what happens in
an atomic event, we have to realize that the word 'happens' can only apply
to the observation , not to the state of affairs between two observations.''
}
\end{quotation}

{\normalsize We may notice that both of the previous citations refer to the
non separability between a subjective element (consciousness in the first,
observation in the second) and the physical world. As we cannot reduce
consciousness to the physical reality underlying it, we cannot describe
quantum events independently from observations. Besides this parallel, we
can mention other suggesting analogies between quantum physics and
consciousness. Let us analyze the following statement made by the
psychologist William James [6]:  }

\begin{quotation}
{\normalsize "Our mental states always have an essential unity, such that
each state of apprehension, however variously compounded, is a single whole
of which every component is, therefore, strictly apprehended (so far as it
is apprehended) as a part."  }
\end{quotation}

{\normalsize If we want to study the physical aspects of a mental state,
classical physics probably would not be appropriate because in general a
classical description is based on the decomposition of a system in a
collection of simple elements which are independent and local. Besides,
every element interacts only with its immediate neighbors [7]. The quantum
description instead is based on a wave function which takes in account all
properties of the system as a whole, and non locality becomes explicit in
the act of measurement. Here non locality means that a measurement on a
spatially localized part of a system may affect instantaneously other
distant parts of it. }

{\normalsize Some people believe that the conscious thought is a non
algorithmic activity, in the sense that it cannot be, in principle,
simulated by a computer. This statement has been presented using
mathematical[8] and philosophical arguments[9]. The mathematical argument is
based on a form of Godel's theorem. The philosophical argument establishes
that if brain activity were algorithmic, then men would not have moral
responsibility for their actions. From the other hand, in quantum physics we
have the property that the result of a single measurement is not computable
from the wave function that describes a given system, because it only gives
information concerning probabilities to obtain any of a set of possible
results. The act of measurement produces what is called the 'collapse' of
the wave function, and the state after this collapse cannot be predicted
deterministically. }

{\normalsize If, with all the arguments presented above, we agree that the
quantum theory is likely to help us in the understanding of consciousness,
we could ask if there are some more specific models of brain functioning
based on it.  }

{\normalsize According to H. P. Stapp [7], an atomic process which is
relevant for the dynamics of brain components is the liberation of
neurotransmitter molecules in the region of the synapses between neurons. If
this process requires a quantum description, then the collection of
processes occurring at all synaptic connections could be described using a
global wave function. At a given time, the wave function will represent a
state which is a superposition of possible outcomes upon observing every
site where these processes occur. If to each of these collections of single
states we associate a macroscopic state of the brain, we could say that at
any time the brain will be in a state that is a superposition of
alternatives. When an appropriate stimulus is presented, one of the
alternatives would be selected, activating what Stapp calls 'top level
events', which would actualize patterns of neural activity in the brain as a
whole. Conscious perceptions are identified here with the feelings of these
top level events.  }

{\normalsize Although the connection between the physical state of the brain
and the experience of consciousness is not fully explained by the model
presented above, we could agree that a quantum approach introduces a non
deterministic element in the flow of conscious thoughts. Sir John Eccles, a
Nobel laureate in Medicine is also aware of this property. He combines his
expertise in neurophysiology with quantum physics to build another
interesting model of consciousness[10]. As Stapp, he starts focusing his
attention in the microscopic processes occurring at the sites of the
synapses. Eccles argues that the uncertainty observed in the generation of
the nerve pulses, associated with the concept of ''dendron'' allows for the
possibility that the actions of a person be influenced by an agent external
to the brain (a non material mind). A dendron is a collection of nerve
fibers which propagate pulses coherently, and its presence in several parts
of the brain seems to be well established. The goal of Eccles is to validate
a dualistic model, according to which mind is non material, independent of
the brain and would interact with it without violating the basic laws of
nature (as energy conservation for example), thanks to the room left by
quantum physics. \vspace{1.2cm}  }

{\normalsize {\large \noindent
CONCLUSIONS } }

{\normalsize {\large \vspace{1cm} }After this brief excursion to the state
of the art on the contributions of physics towards the understanding of
consciousness, we may become motivated to choose between two rather general
approaches to the problem:  }

{\normalsize - physics will bring us closer to the understanding of
consciousness, however, we could never save the barrier imposed by the
presence of certain immaterial agents which take part in the phenomenon
(Eccles).  }

{\normalsize - There is no reason why, some day we will have a full
scientific description of conscious perceptions. This idea may be well
illustrated by the words of Francis Crick:  }

\begin{quotation}
{\normalsize ''Our minds -the behavior of our brains- can be explained by
the interactions of nerve cells (and other cells) and the molecules
associated with them[11].'' }
\end{quotation}

{\normalsize Or we may not feel forced to commit ourselves with an a priori
position. It is very likely that everybody will agree that physics has
contributed, is contributing and will contribute more to the understanding
of consciousness. So, those who have the expertise of this discipline should
be encouraged to dedicate their efforts to solve in part or completely this
challenging problem. \vspace{1.2cm}  }

{\normalsize {\large \noindent
REFERENCES } }

{\normalsize {\large \vspace{1cm} }\noindent
[1] Hodgkin, A. L. and A. F. Huxley, {\it A quantitative description of
membrane current and its application to conduction and excitation in nerve},
J. Physiol. {\bf 117}, 500, 1952.  }

{\normalsize \noindent
[2] Hopfield, J.J., {\it Neural Networks and Physical Systems with Emergent
Computational Abilities}, Proc. Natl. Acad. Sci. U.S.A., {\bf 79}, 2554,
1982.  }

{\normalsize \noindent
[3] Hebb, D. O., {\it The Organization of Behavior}, Wiley, New York, 1949.
}

{\normalsize \noindent
[4] Searle, J. R., {\it The Rediscovery of the Mind}, MIT Press, Cambridge,
MA, 1992, p. 122.  }

{\normalsize \noindent
[5] Heisenberg, W., {\it Physics and Philosophy}, Harper and Row, New York,
1958.  }

{\normalsize \noindent
[6] James, W., {\it The Principles of Psychology}, Dover, New York, 1950, p.
241.  }

{\normalsize \noindent
[7] Stapp, H. P., {\it Mind, Matter and Quantum Mechanics}, Springer Verlag,
Berlin, Heidelberg, 1993.  }

{\normalsize \noindent
[8] Penrose, R., {\it Mathematical Intelligence}, in {\it What is
Intelligence?}, J. Khalfa (ed), Cambridge University Press, 1994, p. 107.  }

{\normalsize \noindent
[9] Bringsjord, S., {\it What Robots can and can't be}, Kluwer, 1992.  }

{\normalsize \noindent
[10] Eccles, J. C., {\it How the Self controls its Brain}, Springer Verlag,
Berlin, Heidelberg, 1994.  }

{\normalsize \noindent
[11] Crick, F., {\it The Astonishing Hypothesis: The Scientific Search for
the Soul}, Simon and Schuster, London, 1994, p.7.  }

\end{document}